# Mathieu-Gauss diffraction of a multi segment-shaped laser beam

## I. Zeylikovich[*] and A. Nikitin

*Physics and Technology department, Yanka Kupala State University of Grodno, 22 Ozheshko street, Grodno, 230023, Belarus*
*[*]zeylikov@gmail.com*

**Abstract:** We studied the field structure and 3-D propagation on diffraction of a multi circular segment-shaped laser beam. Diffraction of a laser beam with the shape of the multi circular segments (two, four and six) was investigated. We have observed for the first time the Mathieu-Gauss (MG)-like modes formation and its propagation in the slow-expansion mode. The fundamental difference between diffraction of the multi-segments' laser beams from the well-known MG beams is the large propagation distance (meters) in the slow-expansion mode without using a focusing lens.

1. Introduction

This study was inspired by the surprising experimental effect such as the development of the beam configuration transmitting in a slow-expansion mode on diffraction of a laser beam with the form of a circular segment [1]. It was thought that diffraction of such beam produces the Mathieu-Gauss (MG) beam-like structure. In this paper, more universal experiment of the laser beam diffraction blocked by the multi segment-shaped aperture was investigated.

Gutiérrez-Vega et al. [2] just proposed an explanation of propagation-invariant optical fields (PIOFs). These fields are well-defined by the Mathieu functions. In [3] the authors first described the experimental observation of Mathieu beams. In this framework, the ideal PIOFs, more recently titled as Helmholtz–Gauss waves in general, have been studied [4]. The author of the paper [5] proposed to use an amplitude-type spatial light modulator to load angular spectrum of Mathieu functions distribution along a narrow annular pupil to produce various forms of Mathieu beams. Bessel-Gauss beams represent a special case of the MG beams when the focal distance of the elliptical coordinates reduces to zero [6]. In [7] the authors considered the MG beam diffraction. It was found that the MG beam diffracts in a slow-expansion (if $\omega_0 < 2/k_t$) or a fast-expansion (if $\omega_0 > 2/k_t$) modes, where $k_t$ is the magnitude of the transverse component of the wave vector, $\omega_0$ is the waist radius of the Gaussian beam. We used converted formulas $\theta_t < \lambda/\pi\omega_0$ (the slow-expansion mode) and $\theta_t > \lambda/\pi\omega_0$ (the fast-expansion mode), where $\theta_t$ is the diffraction angle [1].

In this paper, we report here for the first time the observation of the MG-like modes formation and its propagation in the slow-expansion mode on diffraction of a laser beam with the shape of the multi circular segments (two, four and six). We have explored the light field on diffraction of the multi circular segment-shaped laser beam. The experimental system and results will be defined in the next section. The paper concludes by the discussion and conclusion sections.

2. Experiment

The optical scheme of the experimental setup is shown in Fig.1. The experiment was done by a Gaussian laser beam, a beam diameter of 5 mm, and the wavelength of 0.65 *μm*. In the experiment, laser beams were used consisting of two, four, and six circular segments formed from the initial circular laser beam. As shown in Fig. 1, to obtain two circular segments, a blocking strip (ST, the strip width of 3 mm) with sharp edges glued to a glass plate (GP) was used. To obtain four and six circular segments, square (SQ) and hexagonal (H) blocking plates were used (see Fig. 1). Sharp-edge blocking plates glued to the plane-parallel glass plates were positioned close to the laser exit (L). Diffraction patterns were observed on a white screen (S) located at different distances from the blocking plates and photographed with magnification using a camera (C).

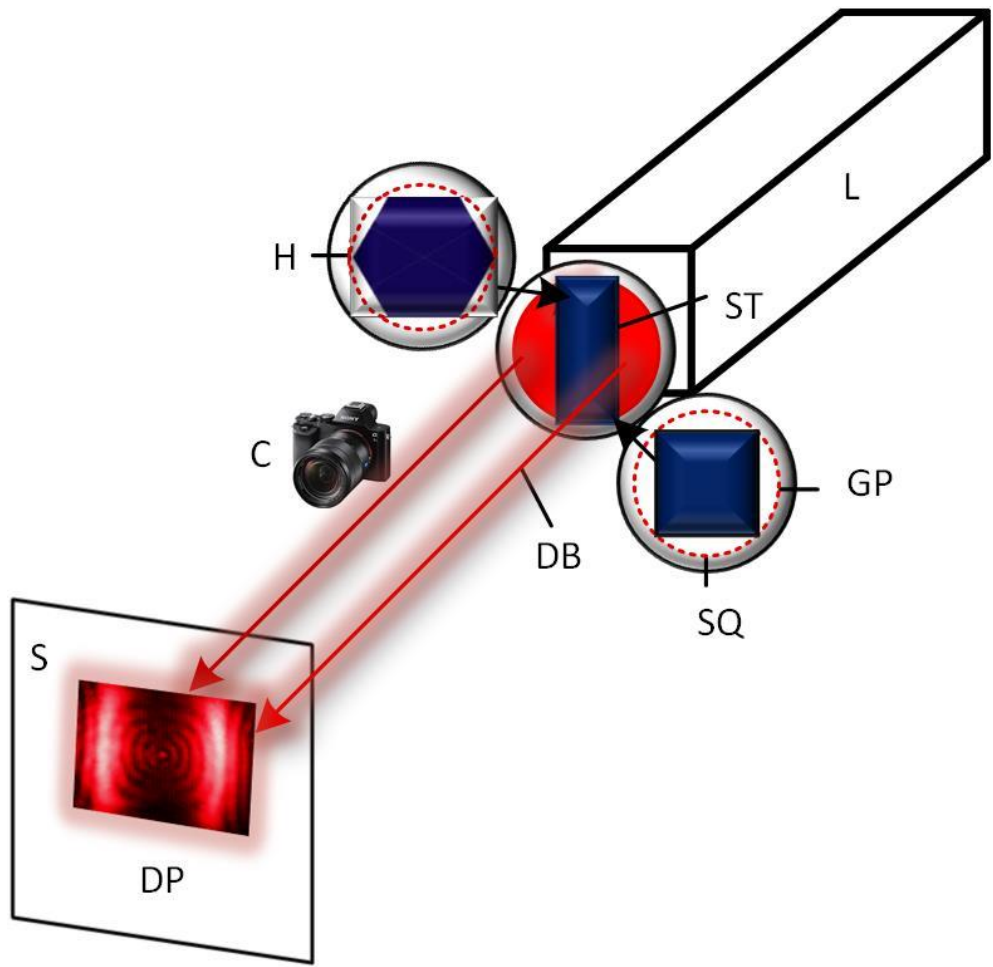

Fig. 1. Experimental setup: L – a laser, ST – a strip, SQ – a square aperture, GP- a glass plate, H – a hexagon aperture, DB – a diffracted beam, S – a white screen, DP – a diffraction pattern, C – a camera.

*2.1. Numerical calculations*

We used the Fresnel-Kirchhoff integral [8, 9] to express the complex amplitude. In calculations $1000 \times 1000$ aperture nodes were used (for the integration) for the aperture diameter of 5 mm. This number of nodes ensure the stability of the solution and the required accuracy. A gradient scale presents the intensity (in relative units) and the phase.

*2.2 Experimental Results*

At the first stage of the experiments, 3-D diffraction of a laser beam formed by two circular segments was studied. The experimental results are presented in Fig. 2 a) -c), and the results of the numerical calculation are shown in Fig. 2 d) -f). The spatial structure (in the XY plane) of the diffraction of the laser beam during propagation along the z-axis was studied at various distances from the blocking aperture. XY-diffraction patterns (a) -c)) were obtained at distances of 150 mm, 200 mm and 300 mm from the blocking strip using the optical scheme shown in Fig. 1. The numerical calculation of (d) -f)), respectively, was performed according to the procedure described in the Section 2.1.

In the following experiments, 3-D diffraction of a laser beam formed by four and six circular segments was studied. The experimental results are presented in Fig. 3 a) -c) and 4 a) -c), respectively, and the results of the numerical calculation are shown in Fig. 3 d) -f) and 4 d) -f), respectively. XY-diffraction patterns (3a) -c)) were obtained at distances of 300 cm, 350 cm and 400 cm from the blocking plate (square), and the patterns shown in Fig. 4 a) -c) were obtained at distances of 330 cm, 410 cm and 490 cm from the blocking plate (hexagon) (according to the optical scheme of Fig. 1). The numerical calculation (3 and 4 d) -f)), respectively, was performed according to the experimental conditions and to the procedure described in Section 2.1.

In figure 2 g), 3 g) and 4 g) the theoretical 3-D intensity distributions along the *x.z*-plane ($y = 0$) at different distances from the blocking apertures are shown. Full width at half maximum (FWHM) of the diffracted beams ($d_p$) at corresponding distances (z) are shown in the table 1.

Table 1

|  | Z, mm | $d_p$, mm | d, mm |
|---|---|---|---|
| A strip | 150 | 0.03 | |
|  | 200 | 0.04 | 0.4 |
|  | 300 | 0.06 | 1.1 |
|  | 400 | 0.07 | 1.8 |
|  | 600 | 0.10 | 3.3 |
|  | 800 | 0.15 | 4.8 |
|  | 1000 | 0.19 | 6.3 |
| A square | 500 | 0.05 | |
|  | 1000 | 0.13 | 2.0 |
|  | 2000 | 0.21 | 6.1 |
|  | 3000 | 0.33 | 10.2 |
|  | 3500 | 0.39 | 12.2 |
|  | 4000 | 0.50 | 14.2 |
| A hexagon | 500 | 0.06 | |
|  | 1000 | 0.23 | 1.6 |
|  | 1500 | 0.30 | 3.2 |
|  | 3300 | 1.00 | 9.1 |
|  | 4100 | 1.93 | 11.6 |
|  | 4900 | 1.26 | 14.2 |

We use the formula $d = \lambda z/\pi\omega_0$ to estimate the central band of the diffraction pattern for distances $z$ shown in the table 1 ($\omega_0 = d_p$ taking at distances of 150 mm and 500 mm, for a strip and a square, a hexagon, respectively). As follows from the above calculations, the multi circular segment-shaped laser beam diffracts in a slow-expansion mode ($d_p \ll d$).

### 2.3 *Theoretical Model*

Figures 5 A)-C) (a) show equivalent theoretical structures of two, four and six segments. To obtain equivalent arrangements, we used the segmentation by the narrow vertical or horizontal stripes. As a result, each segment radiating a diffracted wave can be represented as a combination of a rectangular slit (a virtual slit (VS)) and a narrow radiating circular slit having a complex configuration: with a circular external arc and an internal arc formed by rectangular segments. Figure 5 shows the combinations: A) (a) - two rectangular and 2 circular slits; in B) (a) - four rectangular and 4 circular slits and C) (a) - six rectangular and 6 circular slits. As a result, the diffraction pattern obtained from two segments (Fig. 5 A) (a)) can be considered as a superposition of two diffraction patterns: from a narrow circular aperture separated by a vertical strip, and a diffraction pattern formed by the interference of two vertical slits. Diffraction patterns obtained from four and six slits (Fig. 5 B) (a) and C) (a)) can be considered as superposition of diffraction patterns from a circular aperture separated by four and six absorbing strips and a diffraction pattern formed by the interference of four and six slits, respectively.

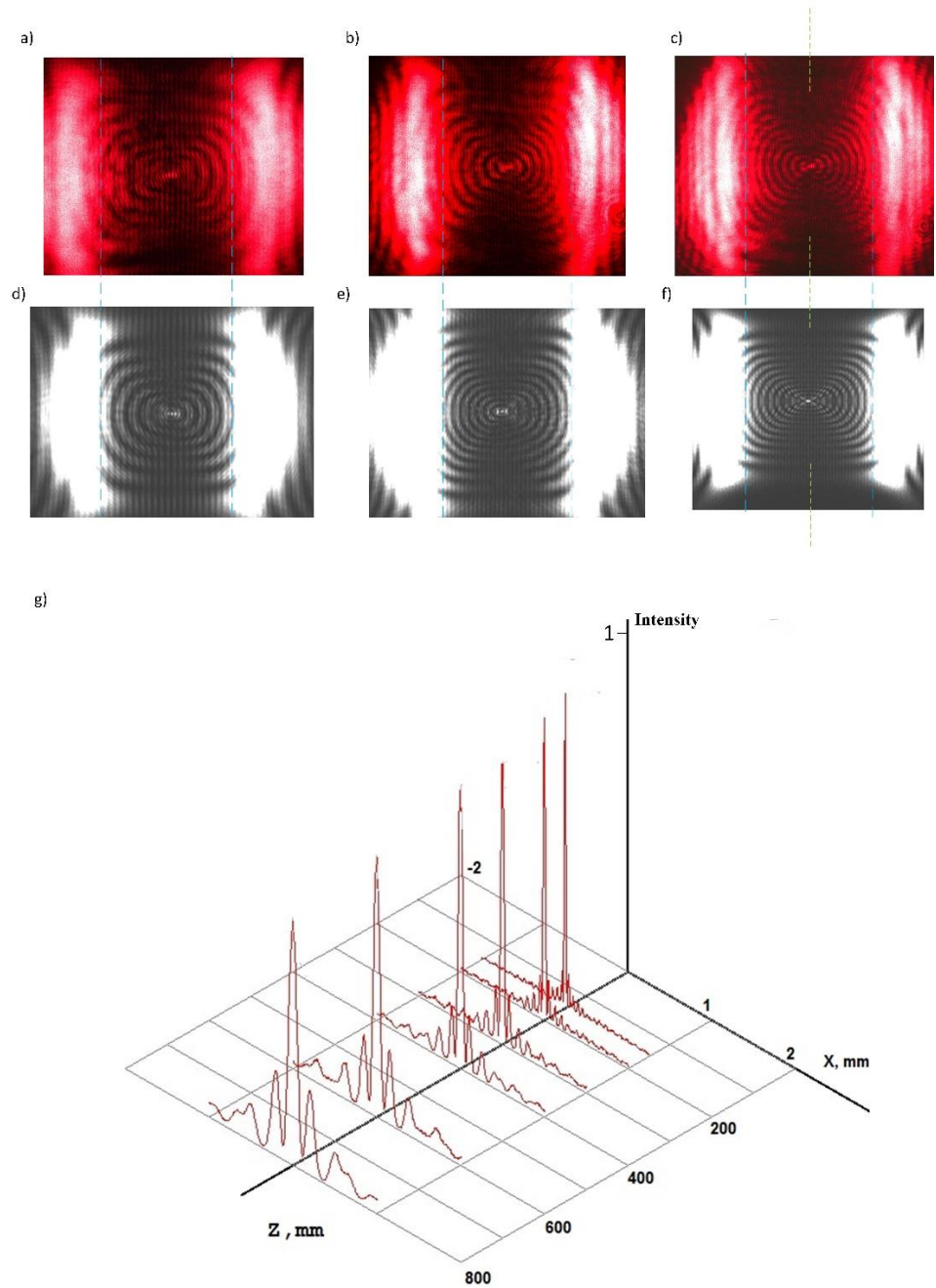

**Fig. 2.** a)-c) are the experimental diffraction patterns (DPs) obtained at distances of 150 mm, 200 mm and 300 mm from the blocking strip, d)-f) are the theoretical DPs obtained by the numerical calculation at the same distances, g) is the theoretical 3-D intensity distribution along the *x.z*-plane ($y = 0$) at different distances from the blocking strip.

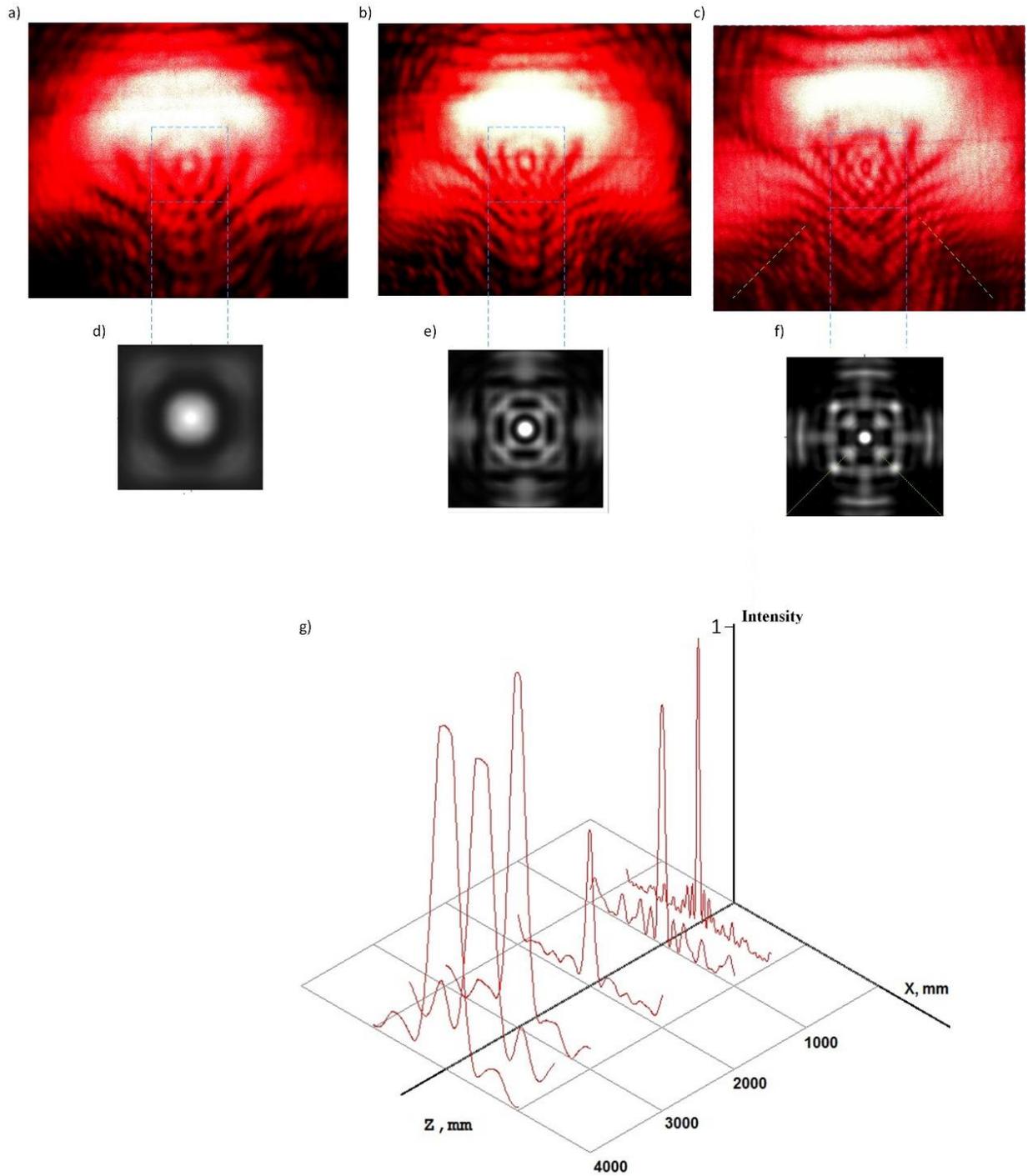

**Fig. 3.** a)-c) are the experimental diffraction patterns (DPs) obtained at distances of 3000 mm, 3500 mm and 4000 mm from the blocking square plate (SP), d)-f) are the theoretical DPs obtained by the numerical calculation at the same distances, g) is the theoretical 3-D intensity distribution along the $x.z$-plane ($y = 0$) at different distances from the blocking SP.

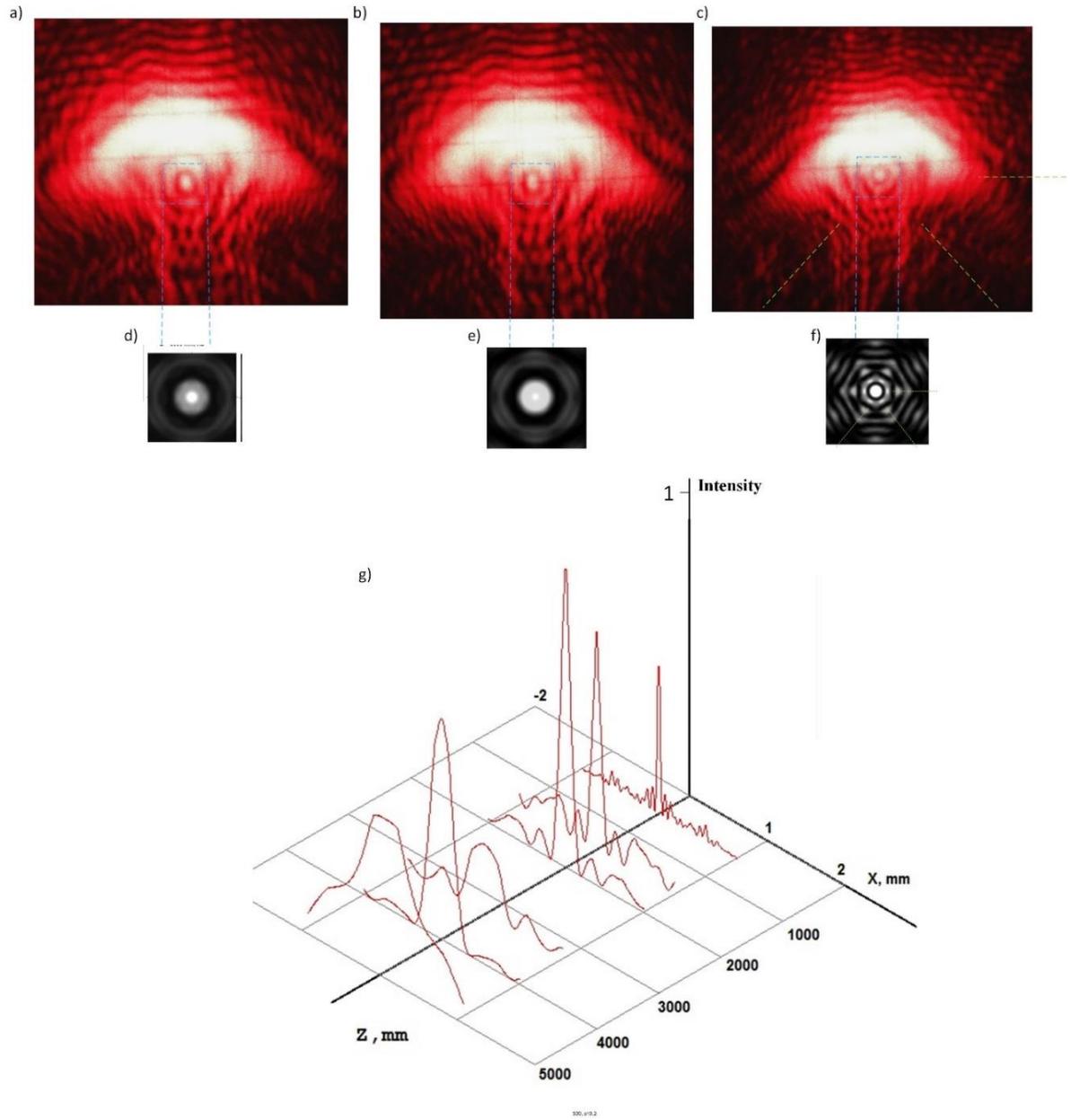

**Fig. 4.** a)-c) are the experimental diffraction patterns (DPs) obtained at distances of 3300 mm, 4100 mm and 4900 mm from the blocking hexagon plate (HP), d)-f) are the theoretical DPs obtained by the numerical calculation at the same distances, g) is the theoretical 3-D intensity distribution along the $x.z$-plane ($y = 0$) at different distances from the blocking HP.

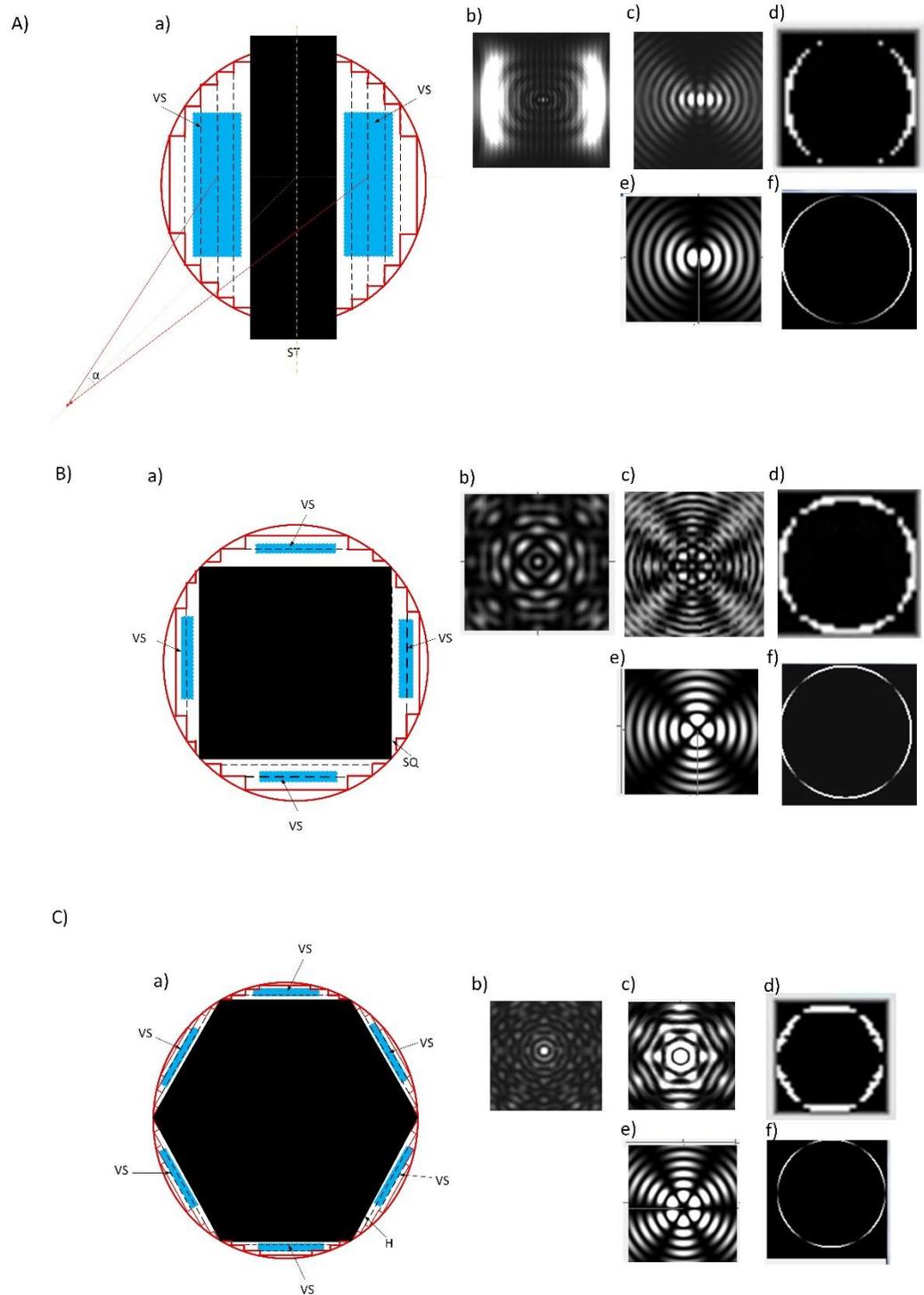

Fig. 5. A)-C) (a) are equivalent theoretical structures of two, four and six segments, VS shows virtual radiating strips, ST, S and H are a blocking strip, square and hexagon apertures. A): b), c) are the theoretical DPs obtained by the numerical calculation at distances of 500, 1000 mm; B): b), c) - 500, 500 mm; C): b), c)-500, 3300 mm. A)-C) d) are narrow virtual radiating slits corresponding to the two, four and six segments of the laser beam. A)-C) e) are the intensity distributions of the Mathieu-Gauss beams (from two, four and six circular slits shown in A) –C) (f)).

Figure 5 A) (b), c)) presents the results of numerical simulations of diffraction patterns formed, respectively, by two segments (b) and 2 radiating circular slits (c), shown in Fig. 5 A) (d). Figure 5 B) (b), c)) shows the outcomes of numerical simulations of diffraction patterns produced, respectively, by four segments and 4 radiating circular slits, shown in Fig. 5 B) (d). Figure 5 C) (b), c)) presents the results of numerical simulations of diffraction patterns formed, respectively, by six segments and 6 radiating circular slits, shown in Fig. 5 C) (d).

To explain these results we apply the solution of the 3-D Helmholtz equation as the product of two functions depending on the z-coordinate and coordinates (x, y) [10, 5]:

$$E(x,y,z) = E_z \cdot E_{xy} \qquad (1)$$

The solution for $E_z$ is trivial:

$$E_z = \exp(ik_z z) \qquad (2)$$

$k_z$- axial wave number;

The equation for $E_{xy}$ in the elliptical-cylindrical coordinate system has the form:

$$\frac{\partial^2 E(\xi,\eta)}{\partial \xi^2} + \frac{\partial^2 E(\xi,\eta)}{\partial \eta^2} + \frac{f^2}{\lambda^2}[ch^2(\xi) - \cos^2(\eta)]E(\xi,\eta) = 0 \qquad (3)$$

where

$E(\xi,\eta)$-an amplitude;

$\xi, \eta, z$- elliptical coordinates;

$f$- half focal length in the elliptical-cylindrical coordinate system;

$\lambda$- wavelength.

The relationship between the Cartesian $x, y, z$ coordinates and the elliptical ones satisfy the relations:

$$x = f \cdot ch(\xi) \cdot \cos(\eta), y = f \cdot sh(\xi) \cdot \sin(\eta), z = z \qquad (4)$$

This form of record (HE) due to the possibility of separation of variables allows us to reduce the problem to the solution of the canonical and modified Mathieu equations. In this case, the solution can be represented as the product of the canonical and modified Mathieu functions [11]:

$$E_r(\xi,\eta) = ce_r(\eta,q) \cdot Je_r(\xi,q) \cdot \exp\left(-\frac{x^2+y^2}{\omega_0^2}\right) \qquad (5)$$

$$E_r(\xi,\eta) = se_r(\eta,q) \cdot Jo_r(\xi,q) \cdot \exp\left(-\frac{x^2+y^2}{\omega_0^2}\right), \qquad (6)$$

where

$q = \frac{f^2 k_t^2}{4}$ – ellipticity factor;

$k_t$- radial wave number;

$\omega_0$- waist radius of a Gauss beam.

$ce_r(\eta,q), Je_r(\xi,q)$- even angular and radial Mathieu functions of order $r$;

$se_r(\eta,q), Jo_r(\xi,q)$- odd angular and radial Mathieu functions of order $r$.

We apply as well the solution of the Helmholtz equation stated by the Whittaker integral [2], explicitly,

$$E(x,y,z) = \exp(ik_z z) \int_0^{2\pi} A(\varphi)\exp[i k_t(x\cos\varphi + y\sin\varphi)]d\varphi, \qquad (7)$$

where $A(\varphi)$ is the angular spectrum of field $E(\boldsymbol{r})$, $k_z$ is the magnitude of the longitudinal component of the wave vector $\boldsymbol{k_0}$, $\varphi$ is an angular variable.

According to [12] the even Mathieu function of order $r$ has the form:

$$ce_r(\eta, q = 0) = \cos(r\eta), \qquad (8)$$

and odd:

$$se_r(\eta, q = 0) = \cos(r\eta - \pi/2) \tag{9}$$

Radial functions $Je_r(\xi, q = 0)$ and $Jo_r(\xi, q = 0)$ we find based on (8) and (9):

$$Je_r(\xi, q = 0) = ce_r(i\xi, q = 0) = \cos(i\xi r) \tag{10}$$

$$Jo_r(\xi, q = 0) = -ise_r(i\xi, q = 0) = -i\cos(ri\xi - \pi/2) \tag{11}$$

Figure 5 A) - C) (e) presents the calculation of the intensity distribution of the Mathieu-Gauss beams shown in Fig. 5 A) -C) (f), using the expressions (5) - (7) and taking into account (8) - (11). The resulting shapes correspond to $r$-order Mathieu-Gauss beams for $r = 2, 3, 4$, respectively. An $r$-order Mathieu beam has $r$ angular nodal lines over the beam center and, when $q = 0$ the radial nodal lines of Mathieu beams are circular [5].

## 3. Discussion

Comparing the patterns shown in Figures 5 A) - C) (c) with the diffraction patterns shown in Figures 5 A) - C) (e), we can conclude that the diffraction patterns (DPs) from two, four, and six segments of the circular laser beam are formed by a superposition of the two DPs. The first DP is produced by the $r$ -order MG-like beams (for $r = 2$, 3, 4) and the second one is formed by interference of radiation from the virtual rectangular slits (two, four and six). It should be noted that the experimentally observed diffraction in the slow-expansion mode is explained by the formation of the MG-like beams on diffraction of the laser beams shaped by two, four, and six circular segments. The superposition of these MG-like beams with diffraction patterns from virtual slits only leads to a redistribution of intensity in diffraction patterns without changing the fundamental property of the MG-beams - propagation in the slow-expansion mode. We notice that the Mathieu-Gauss patterns shown in Fig. 5 e) can be formed in the focal plane of the lens as a result of diffraction of the laser beam by a narrow annular pupil by corresponding modulation of the intensity distribution [3, 5] (with focusing depth of 1-2 cm). The fundamental difference between diffraction of the multi-segments' beams from the well-known Mathieu-Gauss beams is the large propagation length (meters) in the slow-expansion mode without using a focusing lens.

## 4. Conclusion

We report here for the first time the observation of the MG-like modes formation and its propagation in the slow-expansion mode on diffraction of a laser beam with the shape of the multi circular segments. In the experiment, laser beams were used consisting of two, four, and six circular segments formed from the initial circular laser beam. The DPs from two, four, and six segments of the circular laser beam are formed by a superposition of the two DPs. The first DP is produced by the $r$-order MG-like beams (for $r = 2, 3, 4$) and the second one is formed by interference of radiation from the virtual rectangular slits (two, four and six). The major difference between diffraction of the multi-segments' beams from the MG beams is the huge propagation distance (meters) in the slow-expansion mode without using a focusing lens.

Our results can be useful for more deep thoughtful of light diffraction, which may advance the producing of new diffractive optics. A high power laser beam propagating in the slow-expansion mode can be invented.

## 5. Funding, acknowledgments, and disclosures

### 5.1. Funding

This research did not receive any specific grant from funding agencies in the public, commercial, or not-for-profit sectors.

### 5.2 Acknowledgments

### 5.3 Disclosures

The authors declare that there are no conflicts of interest related to this article.

**Author contribution statement**

Both authors discussed the results and contributed to the final manuscript. Iosif Zeylikovich determined the direction of the research, performed the experimental part and theoretical interpretation. Nikitin performed all numerical calculations and participated in the theoretical interpretation.

**References**

1. I. Zeylikovich and A. Nikitin, "The formation of a beam propagating in a slow-expansion mode on diffraction of a circular segment-shaped laser beam," Eur. Phys. J. D (2019) 73: 180. https://doi.org/10.1140/epjd/e2019-90665-3
2. J. C. Gutiérrez-Vega, M. D. Iturbe-Castillo and S. Chávez-Cerda, "Alternative formulation for invariant optical fields: Mathieu beams," Opt. Lett. **251**, 1493-1495 (2000).
3. J. C. Gutiérrez-Vega, M. D. Iturbe-Castillo, G. A. Ramírez, E. Tepichín, R. M. Rodríguez-Dagnino, S. Chávez-Cerda, G. H. C. New, "Experimental demonstration of optical Mathieu beams," Opt. Commun. **195**, 35-40 (2001).
4. J. C. Gutiérrez-Vega and M. A. Bandres, "Helmholtz–Gauss waves," J. Opt. Soc. Am. A **22**, 289-298 (2005).
5. Z. Ren, J. He and Y. Shi, "Generation of Mathieu beams using angular pupil modulation," Chin. Phys. B **27**, 124201 (2018).
6. F. Gori, G. Guattari and C. Padovani, "Bessel-Gauss beams," Opt. Commun. **64**, 491-495 (1987).
7. U. Levy, S. Derevyanko, Y. Silberberg, *"Light modes of free space,"* in Progress in Optics, ed. T. D. Visser, (Elsevier: Amsterdam, 2016), Vol. 61, pp. 237-281.
8. M. Born and E. Wolf, *Principles of Optics* (Oxford: Pergamon Press, 1968).
9. I. Zeylikovich and A. Nikitin, "Diffraction of a Gaussian laser beam by a straight edge leading to the formation of optical vortices and elliptical diffraction fringes,"Opt. Commun. **413**, 261-268 (2018).
10. S. Chávez-Cerda, J. C. Gutiérrez-Vega, G. H. C. New, "Elliptic vortices of electromagnetic wave fields," Opt. Lett. **26**(22), 1803-1805 (2001).
11. J. C. Gutiérrez-Vega and M. A. Bandres, "Elliptic vortices of electromagnetic wave fields," J. Opt. Soc. Am. A **24**, 215-220 (2007).
12. M. Abramovits, I. Stigan (Eds.), *Handbook of Mathematical Functions [in Russian]* (Nauka, Moscow, 1979).